\documentclass[showpacs,showkeys,11pt,
preprint,preprintnumbers,nofootinbib,
groupedaddress,superscriptaddress,amsmath,amssymb]{revtex4}
\usepackage{amsfonts}
\usepackage{graphics}
\usepackage{epsfig}
\usepackage{url}
\usepackage{multirow}
\newcommand {\be}{\begin{equation}}
\newcommand {\ee}{\end{equation}}
\newcommand {\ba}{\begin{eqnarray}}
\newcommand {\ea}{\end{eqnarray}}
\newcommand {\invfb}{$fb^{-1}$}
\newcommand {\tanb}{$\tan\beta~$}
\newcommand {\ra}{\rightarrow}

\begin{document}
\title{Charged Higgs Detection in the $\tau\nu$ Decay Mode at Future Linear Colliders}
\pacs{12.60.Fr, 
      14.80.Fd,  
      14.60.Fg 
}
\keywords{MSSM, Higgs bosons, $\tau$ Lepton, Linear Colliders}
\author{M. Hashemi}
\email{hashemi_mj@shirazu.ac.ir}
\affiliation{Physics Department and Biruni Observatory, College of Sciences, Shiraz University, Shiraz 71454, Iran}

\begin{abstract}
Charged Higgs phenomenology and discovery potential at future linear colliders through the production process $e^{-}e^{+} \ra H^{+}H^{-} \ra \tau^{+}\nu \tau \bar{\nu}$ is studied. Both charged Higgs bosons are considered to decay to $\tau\nu$ and the hadronic decay of $\tau$ leptons is analyzed taking into account spin and kinematics effects using a proper simulation of the $\tau$ lepton decay. It is shown that within the MSSM framework, with \tanb = 10, a wide range of charged Higgs masses would have detectable signal beyond the $5\sigma$ statistical significance at linear colliders with $\sqrt{s} = 500$ and $1000$ GeV.   
\end{abstract}

\maketitle

\section{Introduction}
The Standard Model of particle physics (SM) has acquired a magnificent success in describing sub-atomic events and particles properties in the last years. The Higgs mechanism is now thought to be the correct underlying mechanism for giving mass to the elementary particles. Signals of a boson with a mass around 125 GeV has recently been observed at the LHC \cite{125atlas,125cms} and there is strong belief that this signal is really the Higgs boson. While a large part of effort has been in the direction of SM Higgs boson searches, theories beyond SM are also under attention. In supersymmetric models beyond SM \cite{Martin}, advantages appear: the quadratic divergence of the Higgs boson mass is removed naturally by including supersymmetric particle contributions in the Feynman diagrams, gauge unification is achieved and candidates for the dark matter are also proposed. \\
In the so called Two Higgs Doublet Model, 2HDM, more than a single Higgs boson is predicted. The minimal supersymmetric standard model (MSSM), belonging to 2HDMs family, expects five Higgs bosons, two of which are charged. While neutral MSSM Higgs bosons may appear similar to their SM partner, existence of a charged Higgs boson would be a signature of models beyond SM. \\
The charged Higgs boson has been searched for extensively in recent high energy experiments. The search strategies are based on direct and indirect searches. In direct searches, a charged Higgs signal is searched for through production processes which involve this particle and its decay products. The search is therefore based on an excess of events over what is expected from SM processes withought charged Higgs. The current results of such searches include the lower limit set by the LEP Higgs Working Group which excludes a charged Higgs with $m_{H^{+}}<80$ GeV \cite{lepexclusion1}. The CDF collaboration has also excluded high \tanb region of parameter space \cite{cdfexclusion}. The current result from the CMS collaboration at LHC excludes a wider region of $(m_{H^{\pm}},\tan\beta)$ space compared to previous experiments \cite{CHCMS}. Based on this result, \tanb= 10 can still be used for all charged Higgs masses.\\ 
The indirect search strategies are based on observation of any deviation from SM, which arises when Feynman diagrams with charged Higgs propagators are added in the calculation. The result appears as slight changes in cross sections, branching ratio of decays of known particles or other observables compared to SM withought Higgs. One of such analyses searches for SM $t\bar{t}$ cross section deviations from what is expected in different final states \cite{D0Indirect}. The existence of a charged Higgs is then inferred as imbalance between cross section of different final states when corrected for lepton identification efficiencies. A review of direct and indirect searches at the Tevatron can be found in \cite{TeVReview}. Results from B-meson decays can also impose indirect constraints on the charged Higgs mass. The strongest limit arises from the partonic transition $b\ra s\gamma$ which excludes a charged Higgs with mass below 295 GeV at 95 $\%$ C.L. in 2HDM Type II for \tanb higher than 2 using CLEO data \cite{B1}.
However it is not obvious how to translate results of B-physics studies to a supersymmetric 2HDM like MSSM which is the framework of this work. Therefore in this paper the charged Higgs direct search results are taken as the bottom line.\\
In a different way, limits on the mass of other neutral Higgs bosons can be translated to limits on the charged Higgs mass through the relations which hold between their masses. An analysis of this kind performed by the LEP Higgs working group, uses the combined result of MSSM neutral Higgs boson searches and excludes a light charged Higgs with $m_{H^{+}} < 125$ GeV \cite{lepexclusion2}.\\
While the charged Higgs is under attention in the current experiments, one may face situations, in which, possible observation of a charged Higgs is postponed to the future colliders. Suppose LHC observes a single light neutral Higgs boson which can also be interpreted as the lightest neutral MSSM Higgs boson, $h^{0}$. In the so called ``decoupling limit'', i.e., when  $m_{A^{0}}\gg m_{Z^{0}}$, the lightest MSSM Higgs boson, $h^{0}$, reaches the upper limit $m^{2}_{h^{0}}\approx m^{2}_{Z^{0}}\cos^{2}2\beta~+$ loop corrections, while the other Higgs bosons, $A^{0}$, $H^{0}$ and $H^{\pm}$ would be much heavy and nearly degenerate through their mass relations as in Eq. \ref{masses}. 
\be
m^{2}_{h^{0},H^{0}}=\frac{1}{2}\left(m^{2}_{A^{0}}+m^{2}_{Z^{0}}\mp\sqrt{(m^{2}_{A^{0}}-m^{2}_{Z^{0}})^{2}+4m^{2}_{Z^{0}}m^{2}_{A^{0}}\sin^{2}(2\beta)}\right), ~~~ m^{2}_{H^{\pm}}=m^{2}_{A^{0}}+m^{2}_{W}
\label{masses}
\ee
In this case $m_{H^{0}}\approx m_{H^{\pm}} \approx m_{A^{0}}$. Therefore the charged Higgs boson may be ``decoupled'' from the current experiments and only be observed at a future collider. \\

\begin{figure}[h]
\begin{center}
\includegraphics[width=0.6\textwidth]{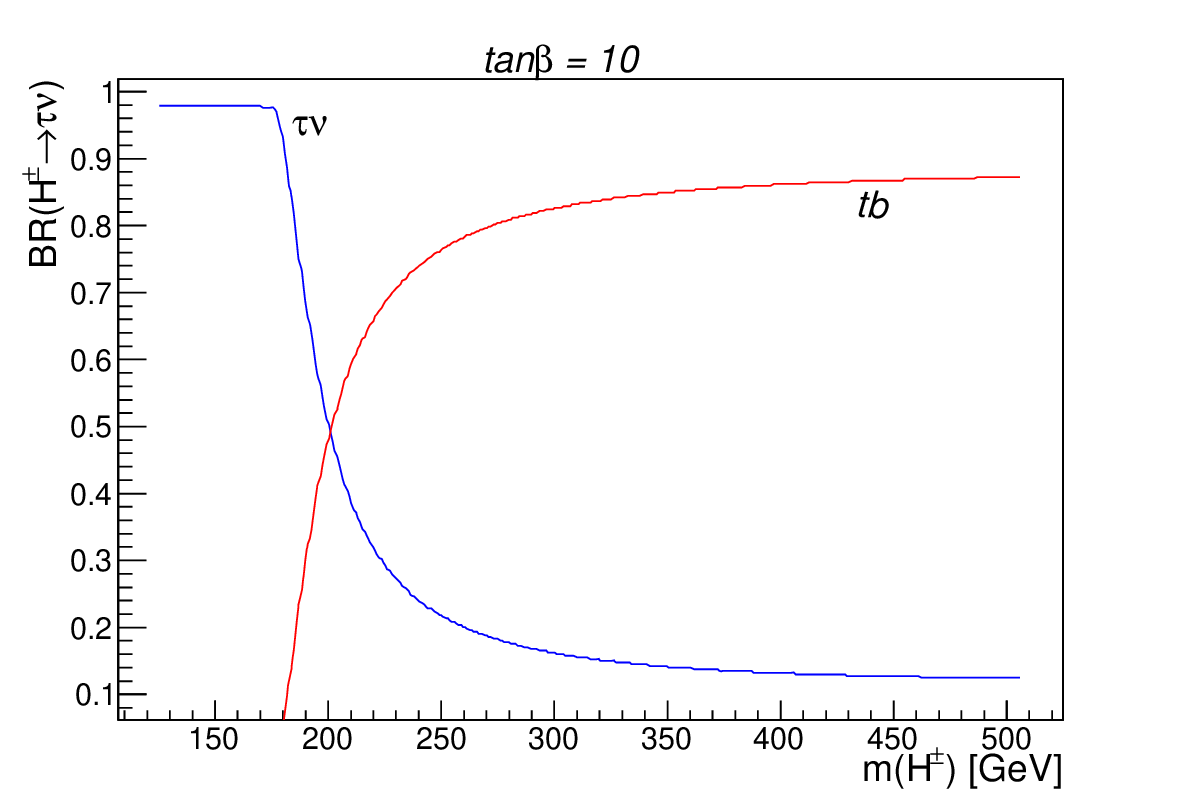}
\end{center}
\caption{Branching ratio of charged Higgs decay to different particles as a function of its mass.}
\label{brch}
\end{figure}

In a different direction, suppose \tanb is small. The charged Higgs couplings are listed in Eq. \ref{lag}. The $H^{+}\bar{t}b$ coupling decreases with decreasing \tanb as long as $\tan\beta > \sqrt{m_{t}/m_{b}}\simeq 6$. This can be easily verified by plotting $m_{t}\cot\beta+m_{b}\tan\beta$ as a function of \tanb \cite{LCHCMS}. Of course, the above coupling starts to behave in the opposite way if \tanb is decreased further, however, these points are very close or inside the excluded area of LEP \cite{lepexclusion2} and are not considered here. The case of $H^{+}\tau\nu$ always remains proportional to \tanb and again decreases with decreasing \tanb.  
\be
H^{+}\bar{t}b : \frac{g}{\sqrt{2}M_{W}}(m_{t}\cot\beta+m_{b}\tan\beta),~~~ H^{+}\tau\nu : \frac{g}{\sqrt{2}M_{W}}m_{\tau}\tan\beta 
\label{lag}
\ee
Therefore two scenarios of light and heavy charged Higgs arise in this case as the following. If the charged Higgs is light enough to be produced in a top quark decay, i.e., $t\ra H^{+}b$, then it decays predominantly to a $\tau\nu$ pair. Therefore the main production process would be a top pair produced in proton-proton collisions at LHC, followed by the top quark decay to charged Higgs which decays subsequently to $\tau\nu$ \cite{LCHCMS}. Such a process involves $H^{+}\bar{t}b$ and $H^{+}\tau\nu$ vertices. The strength of these vertices decreases rapidly at low \tanb values. In the heavy charged Higgs scenario, one may consider the charged Higgs produced through $gg\ra t\bar{b}H^{-}$ and $gb \ra tH^{-}$ described in \cite{TPlehn,TPlehn2}, followed by its decay to $\tau\nu$ \cite{HCHCMS} or $t\bar{b}$ \cite{LowetteCH}. In both cases the production process again involves $H^{+}\tau\nu$ or $H^{+}\bar{t}b$ vertices through the charged Higgs decay and a similar argument as above applies. As the result, it would be hard for LHC experiments to probe for charged Higgs production processes if \tanb is small. \\
The above arguments imply that at a future linear collider, the charged Higgs studies would be of interest, especially if LHC fails to confirm or exclude the existence of this particle due to the high $m_{H^+}$ or low \tanb. These two domains are expected to be better explored in a linear lepton collider. In this paper, focus is on a linear $e^{-}e^{+}$ collider operating at a center of mass energy of 500 or 1000 GeV. Such a collider may be the International Linear Collider (ILC) \cite{ilc,rdr} or the Compact Linear Collider (CLIC) \cite{clic} running in its low energy phase. As will be seen, a heavy charged Higgs (although not really in the decoupling limit) is observable with \tanb as low as 10; a point which is hardly accessible at the LHC \cite{CMScontour}.\\
There has been an extensive work on estimating the discovery potential of a charged Higgs boson in $e^{+}e^{-}$ linear colliders. Such colliders possess the potential of charged Higgs observation beyond the LHC reach. A charged Higgs lighter than the top quark has been studied in \cite{CH1ILC} with an emphasis on the $\tau$ lepton polarization effects. The pair production, $e^{+}e^{-}\ra H^{+}H^{-}$, has been studied in \cite{pair}, however since it is limited to $m_{H^{\pm}}\leq \sqrt{s}/2$, the single charged Higgs production processes have attracted interest in the literature. In \cite{sch1,sch2,sch3}, the single charged Higgs production through different processes was analyzed. These processes can probe areas of the parameter space not accessible by the pair production process. The $e^{+}e^{-}\ra t\bar{b} H^{-}$ was studied in \cite{sch4,sch5}. However it was shown that the cross section of this process is a fraction of femtobarn leading to no hope for observation of this channel. In \cite{sch6,sch7}, the single heavy charged Higgs production through $e^{+}e^{-}\ra \tau\bar{\nu} H^{+}$ was analyzed. It was shown that the charged Higgs pair production is increased when including off-shell effects and leads to more promising results. Since then more attention has been paid on single charged Higgs production when a heavy charged Higgs is the target. However results from \cite{sch6,sch7} show that including off-shell effects, the $5\sigma$ contour would be extended by only about 10 GeV compared to the case of on-shell charged Higgs pair production. The $e\nu_{e}H^{\pm}$ process was also studied in \cite{sch8}, but it was concluded that the signal cross section is below 0.01 $fb$ even for the very low \tanb values. The $W^{\pm}H^{\mp}$ cross section was calculated in \cite{WH1,WH2} and possible enhancement of its cross section was examined by including quark and Higgs-loop effects in \cite{WH3}. This is a process which could be of relevance at low \tanb values, however, it was concluded in \cite{WH4,WH5} that only few events of this kind may be observed at $e^{+}e^{-}$ linear colliders. Therefore there is little hope for $W^{\pm}H^{\mp}$ process to be detectable at $e^{+}e^{-}$ colliders, although it may be detectable at high \tanb values at a muon collider \cite{mumuWH}. 
Concluding the above introduction, the single charged Higgs production through $e^{+}e^{-}\ra \tau\bar{\nu} H^{+}$ has been proved to be the most promising channel to search for, in the high mass region while the analogous process, i.e., the pair production of charged Higgs bosons would be the best channel for a light charged Higgs below the kinematic limit $\sqrt{s}/2$. It is reasonable to use the charged Higgs decay to top quarks when a heavy charged Higgs is being analyzed and that is in fact was has been done in \cite{sch6,sch7}. The charged Higgs branching ratio of decay to top quarks starts to be dominant when the charged Higgs mass goes beyond that of the top quark, allowing the charged Higgs boson to decay to a real on-shell top quark. However it is interesting to study the charged Higgs decay to $\tau\nu$ pair as it is the second decay channel for the heavy charged Higgs region. Below the top quark threshold, this decay channel is the dominant one. Figure \ref{brch} compares branching ratio of charged Higgs decay to $\tau\nu$ and $t\bar{b}$, calculated with HDECAY 3.4 \cite{hdecay}. As seen from Fig. \ref{brch} the charged Higgs decay to $\tau\nu$ decreases to the level of roughly 0.2 for the high mass region. This fact reduces the total event rate and signal observability compared to the case of charged Higgs decay to $t\bar{b}$. However combining the analysis results of the two decay modes in a proper statistical method may increase the machine sensitivity to the charged Higgs signal. An example of such a combination, although performed for the SM Higgs boson searches at LHC, can be found in \cite{higgscombination}. Concluding the above summary of previous results, the most competitive search channel to the study presented in this work is the one studied in \cite{sch6,sch7} whose results are compared in detail with those of this study in a separated section. The other search channels do not provide a comparable result to that of \cite{sch6,sch7} nor this study.\\
In the following a numerical analysis of charged Higgs decay to $\tau\nu$ through the production process $e^{+}e^{-} \ra H^{+}H^{-} \ra \tau^{+} \nu \tau \bar{\nu}$ is performed. Therefore the analysis is based on the on-shell production of charged Higgs bosons and the possible off-shell production is not considered here as it may not change the results very much. This is an expectation inspired from \cite{sch6,sch7}.\\ 
The organization of the paper is as the following. In the next section, event generation is described and technical tools used in the analysis are introduced. Section III is devoted to the signal and background processes and their cross sections calculation. In section IV, the analysis strategy is presented and selection cuts are applied to increase the signal to background ratio. Finally in section V, results are presented in terms of selection efficiencies listed in tables and the signal significance as a function of the charged Higgs mass. The paper ends with conclusions on the observability of the charged Higgs in the studied channel.\\

\section{Event Generation}
For the simulation of the signal and background events, and cross section calculations, PYTHIA 8.1.53 \cite{pythia} is used. The SUSY-HIT package \cite{susyhit} is used as a self-contained tool for the calculation of the MSSM Higgs boson and SUSY particle decays. The particle decays are calculated by HDECAY \cite{hdecay} and SDECAY \cite{sdecay} which are both included in SUSY-HIT. For the calculation of the particle spectrum, the renormalization group evolution program SuSpect \cite{suspect} is used. This program is linked to the SUSY-HIT package by default. The output including the particles mass spectra and decays is written in SLHA format \cite{slha} and used by PYTHIA for event generation. \\
The two $\tau$ leptons in the final state are allowed to decay to all possible final states, therefore, no branching ratio of $\tau$ hadronic decay in event calculations is used, however, the analysis is designed to select only the hadronic decay. For a proper simulation of $\tau$ lepton decay, TAUOLA C++ interface \cite{tauola} is linked to PYTHIA. This package is a C++ version of the Fortran-based TAUOLA \cite{tauola1,tauola2,tauola3} and has been designed to be used by PYTHIA 8 series. The output of the PYTHIA is translated to HEPMC 2.05.01 format \cite{hepmc} and is transferred to TAUOLA interface to add the $\tau$ lepton decay information to the event. Having generated events, the jet-like $\tau$ hadronic decays are identified using FASTJET 2.4.1 \cite{fastjet} which is a jet reconstruction package. The anti-kt algorithm \cite{antikt} and a cone size of 0.4 and the ET recombination scheme are used for the jet reconstruction.\\
Finally when events are generated, kinematic distributions are visualized and analyzed using ROOT 5.30 \cite{root}.

\section{The Signal and Background Processes and their Cross Sections}
\begin{figure}
\begin{center}
\includegraphics[width=0.6\textwidth]{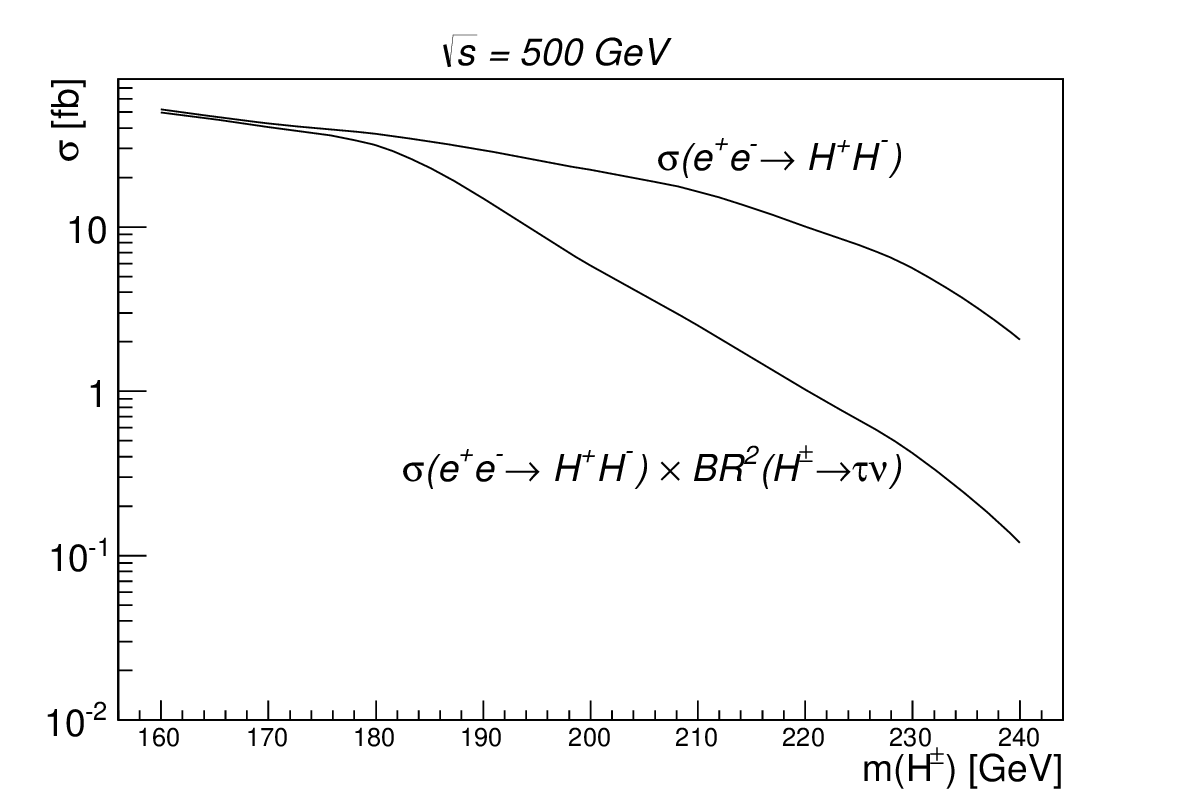}
\end{center}
\caption{The signal cross section, $\sigma$ and $\sigma \times BR$ for a center of mass energy of 500 GeV.}
\label{xsec1}
\end{figure}

\begin{figure}
\begin{center}
\includegraphics[width=0.6\textwidth]{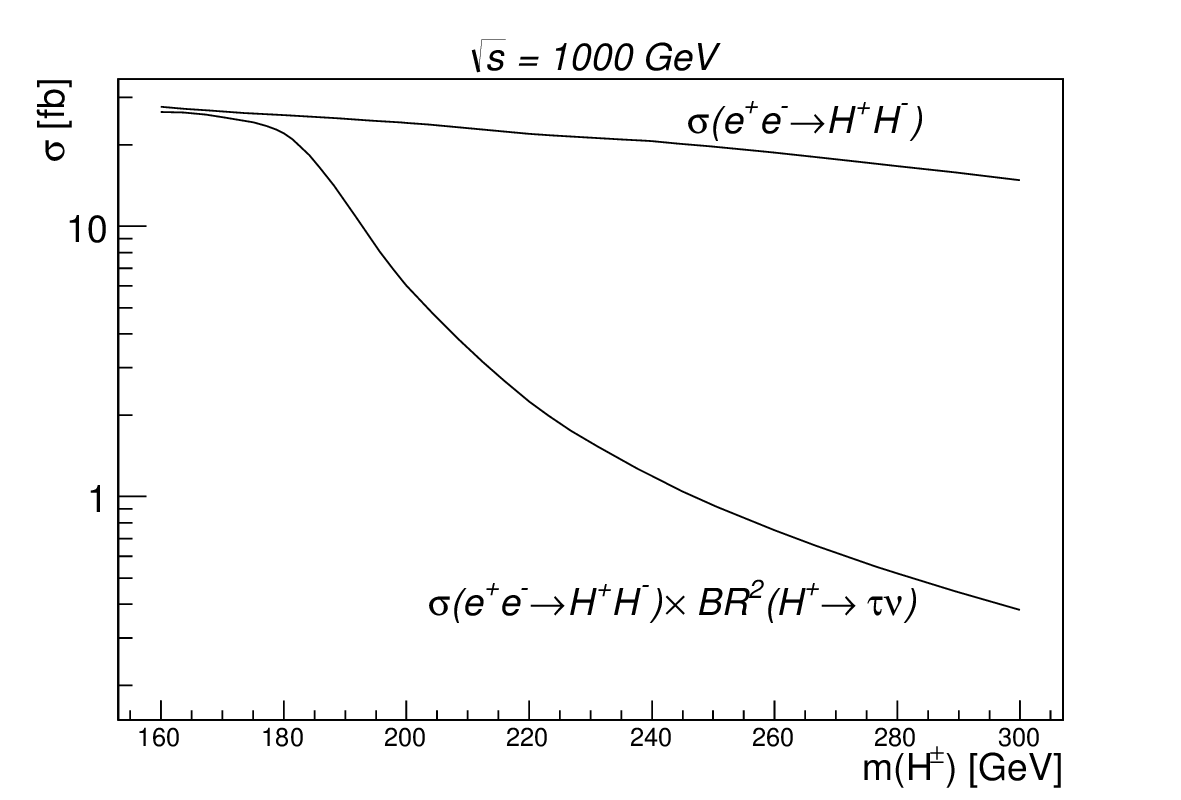}
\end{center}
\caption{The signal cross section, $\sigma$ and $\sigma \times BR$ for a center of mass energy of 1000 GeV.}
\label{xsec2}
\end{figure}

As discussed before the signal process is:\\
\be
e^{+}e^{-} \ra H^{+}H^{-} \ra \tau^{+}\nu\tau\bar{\nu}
\ee
which proceeds mainly through an $s$-channel diagrams with $\gamma,~Z^0,~h^0,~H^0$ and $A^0$ being involved in the propagator. The $t$-channel diagram which involves exchange of a neutrino has a small contribution due to the small coupling of the charged Higgs and electron (or positron), however all diagrams are taken into account in the simulation. The main background processes for this signal would be Drell-Yan process,\\
\be
e^{+}e^{-} \ra Z/\gamma^{*} \ra \tau^{+}\tau 
\ee 
the pair production of $Z$ bosons,\\
\be
e^{+}e^{-} \ra ZZ \ra \tau^{+}\tau\nu\bar{\nu}
\ee 
and $W$ boson pair production,
\be
e^{+}e^{-} \ra W^{+}W^{-} \ra \tau^{+}\nu\tau\bar{\nu}
\ee 
The $Z/\gamma^{*}$ process contains some missing $E_{T}$ due to the $\tau$ lepton hadronic decay which produces $\tau$ neutrinos. Any imbalance of energy as a result of mid-identification of jets can also result in fake missing $E_{T}$ in the event. Therefore such events may appear with the same final state as the signal and as will be seen later, they contribute as the main source of the background. The $ZZ$ process has a small cross section and few tens of such events survive the event selection as will be seen in the results section. \\
It should be noted that when searching for events with $\tau^{+}\tau E^{miss}_{T}$ as the final state, other sources of the charged Higgs production may also contribute to the signal. An example of such events would be $e^{+}e^{-} \ra W^{\pm}H^{\mp}$. These events have a negligible cross section at high \tanb. As a comparison, with \tanb = 10 and $m_{H^{\pm}}~=~160$ GeV, the cross section of $e^{+}e^{-} \ra W^{\pm}H^{\mp}$ is about $0.1~fb$ \cite{WH3}, which is much smaller than the corresponding cross section of $e^{+}e^{-} \ra H^{+}H^{-}$ ($\sim 52~fb$) for the same values of \tanb and $m_{H^{\pm}}$. Therefore in the search for the charged Higgs boson using the pair production channel, any contribution to the signal due to $W^{\pm}H^{\mp}$ would be a small fraction of percent. \\  
Concerning the theoretical framework, the MSSM is adopted for this search. For the signal simulation the LEP $m_{h}-max$ benchmark scenario is used with the following parameters: $M_{2}=200$ GeV, $M_{\tilde{g}}=800$ GeV, $\mu=200$ GeV and $M_{SUSY}=1$ TeV. The $m_{h}-max$ scenario has been defined to yield the maximal value of $m_{h}$, thus minimizing the excluded area in the parameter space. This scenario is used to set the most conservative exclusion bounds on the MSSM Higgs boson masses and \tanb for fixed values of the top quark mass and $M_{SUSY}$ \cite{lepexclusion2}. Therefore it is adopted as the working point in this study. The above set of parameters yield a light SM-like neutral Higgs ($h$) in the range 123 GeV $<m_h<$ 128 GeV with 160 GeV $<m_{H^{\pm}}<$ 300 GeV and \tanb= 10 using FeynHiggs 2.8.3 two loop level calculations \cite{fh1,fh2,fh3,fh4}. Therefore the scenario is in agreement with LHC observation of a light neutral Higgs boson \cite{125atlas,125cms}. As stated before, \tanb = 10 is used throughout the paper. This is a value currently outside the excluded area in \cite{CHCMS}. In order to calculate the signal cross section, the charged Higgs branching ratio of decay to $\tau\nu$ is taken into account and the total cross section, $\sigma$, and the cross section times branching ratio, $\sigma \times BR$, are calculated for two different scenarios of $\sqrt{s}=500$ and 1000 GeV. Results are presented in Figs. \ref{xsec1} and \ref{xsec2}. \\
A linear collider with $\sqrt{s}=1000$ GeV may be expected to perform better in the heavy charged Higgs mass regions not accessible by a collider operating at $\sqrt{s}=500$ GeV. In such regions, the branching ratio of charged Higgs decay to $\tau \nu$ and thus $\sigma \times BR$ are small, as can be seen from Fig. \ref{xsec2}. Therefore the machine sensitivity to the heavy charged Higgs signal in the final state studied in this analysis may be poor even for a collider with $\sqrt{s}$ = 1000 GeV. In the following the analysis is described in details based on $\sqrt{s}$ = 500 GeV, while at the end, the whole analysis is repeated with $\sqrt{s}$ = 1000 GeV and final results are presented and compared with those obtained with $\sqrt{s}$ = 500 GeV.

\section{Event Selection and Analysis Strategy}
\begin{figure}
\begin{center}
\includegraphics[width=0.6\textwidth]{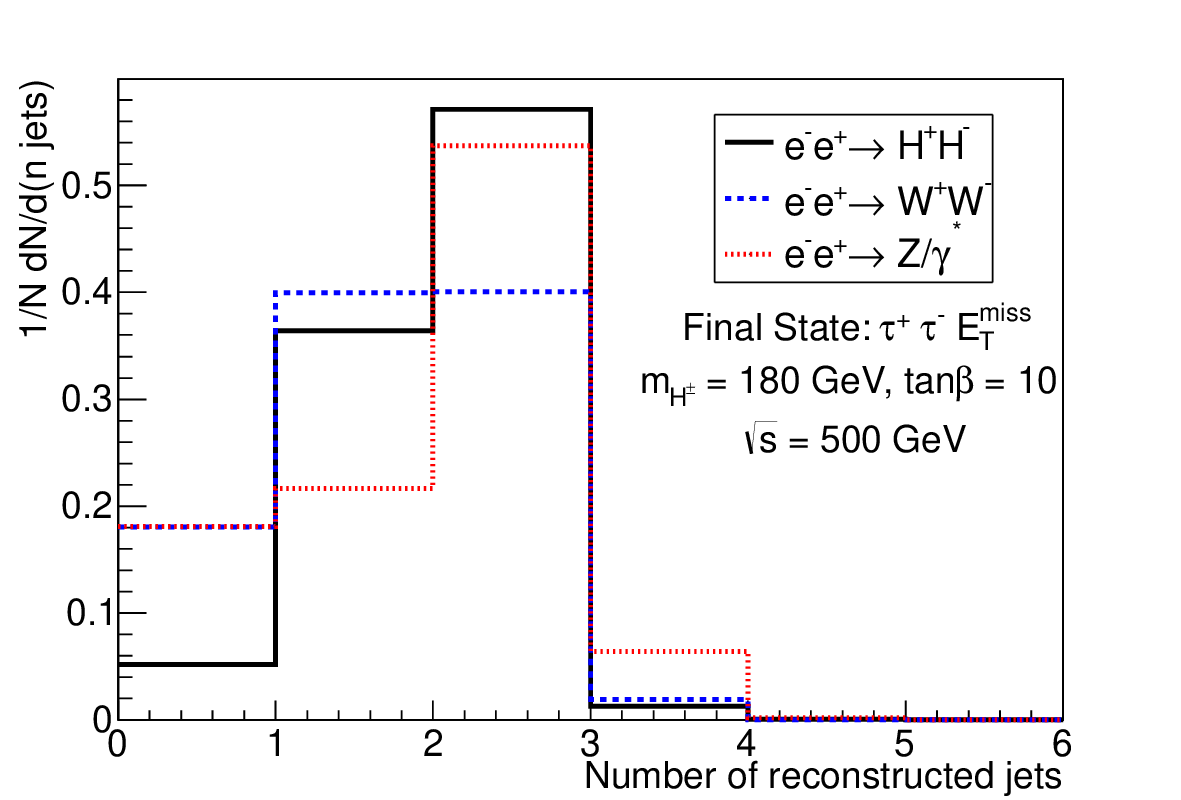}
\end{center}
\caption{The jet multiplicity in signal and background events for a center of mass energy of 500 GeV.}
\label{njets}
\end{figure}

\begin{figure}
\begin{center}
\includegraphics[width=0.6\textwidth]{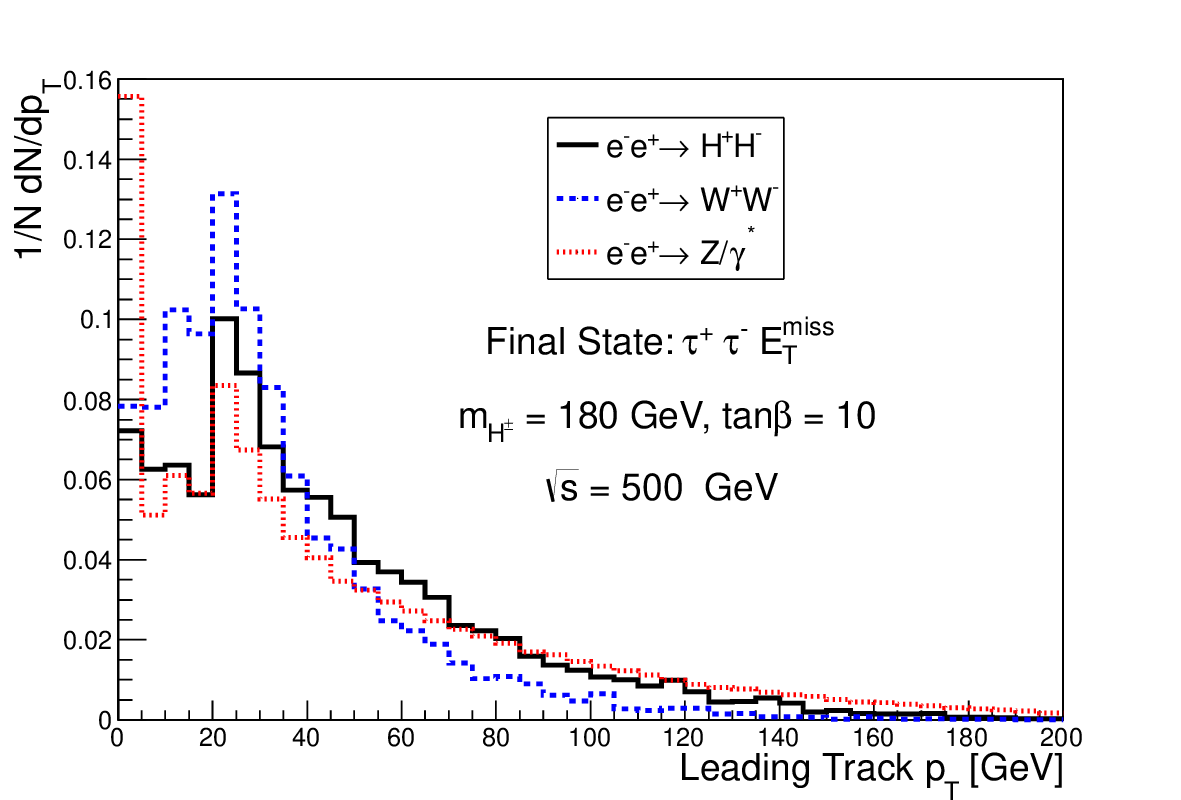}
\end{center}
\caption{The leading track $p_{T}$ distribution in signal and background events for a center of mass energy of 500 GeV.}
\label{ltk}
\end{figure}

\begin{figure}
\begin{center}
\includegraphics[width=0.6\textwidth]{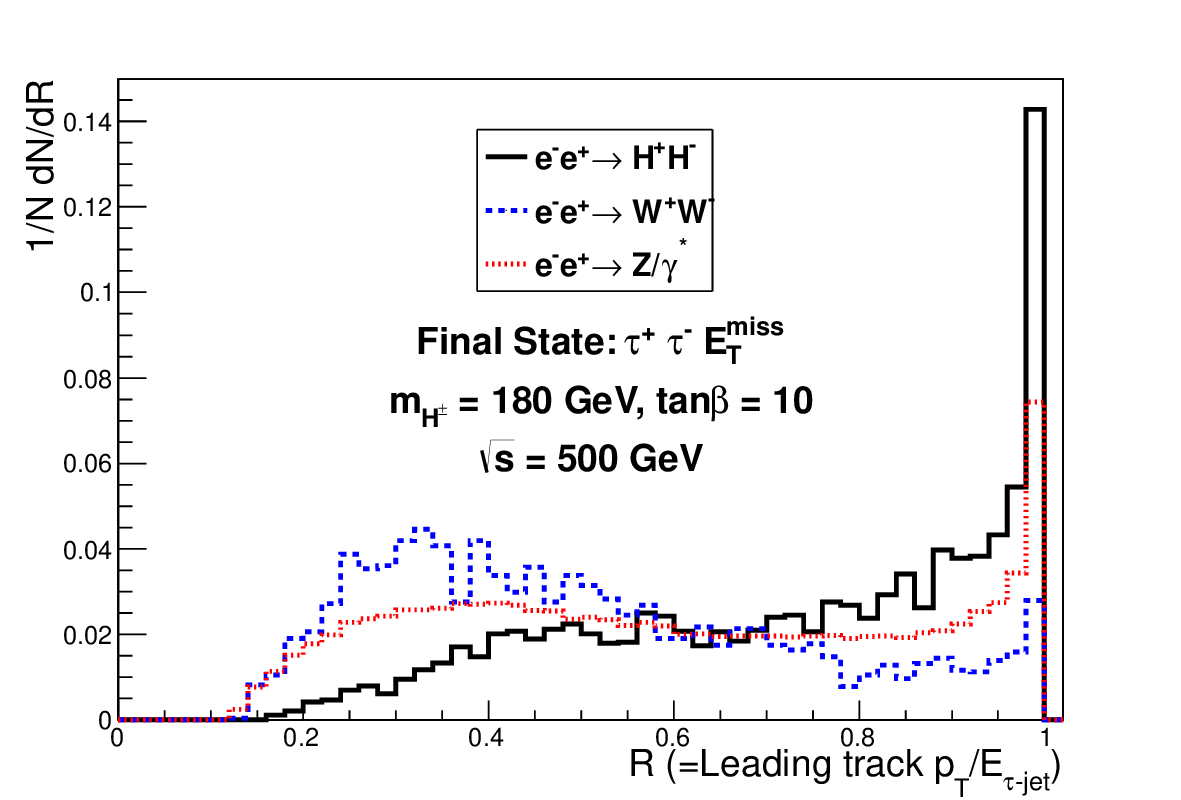}
\end{center}
\caption{Distribution of the leading track $p_{T}$ divided by the $\tau$ jet energy denoted as $R$. Both signal and background distributions are shown for a center of mass energy of 500 GeV.}
\label{R}
\end{figure}

\begin{figure}
\begin{center}
\includegraphics[width=0.6\textwidth]{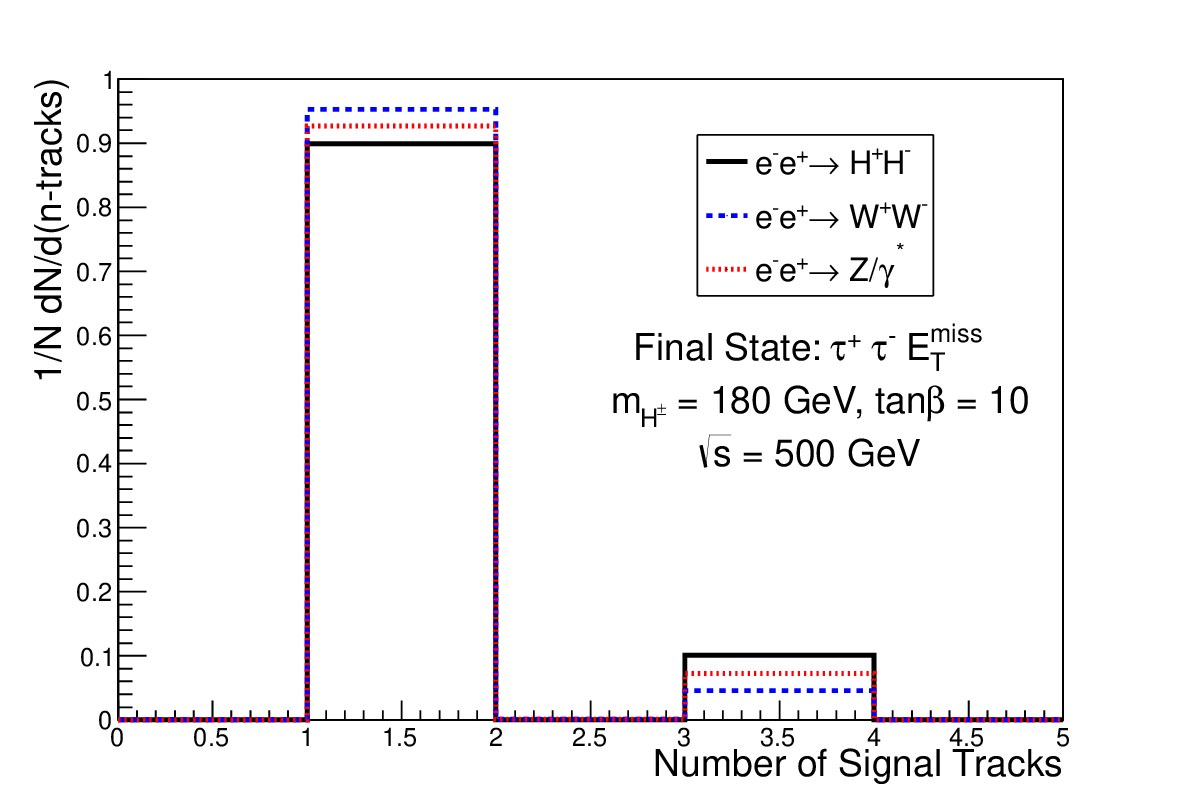}
\end{center}
\caption{Number of signal tracks in the $\tau$ lepton decay in signal and background events for a center of mass energy of 500 GeV.}
\label{nstk}
\end{figure}

\begin{figure}
\begin{center}
\includegraphics[width=0.6\textwidth]{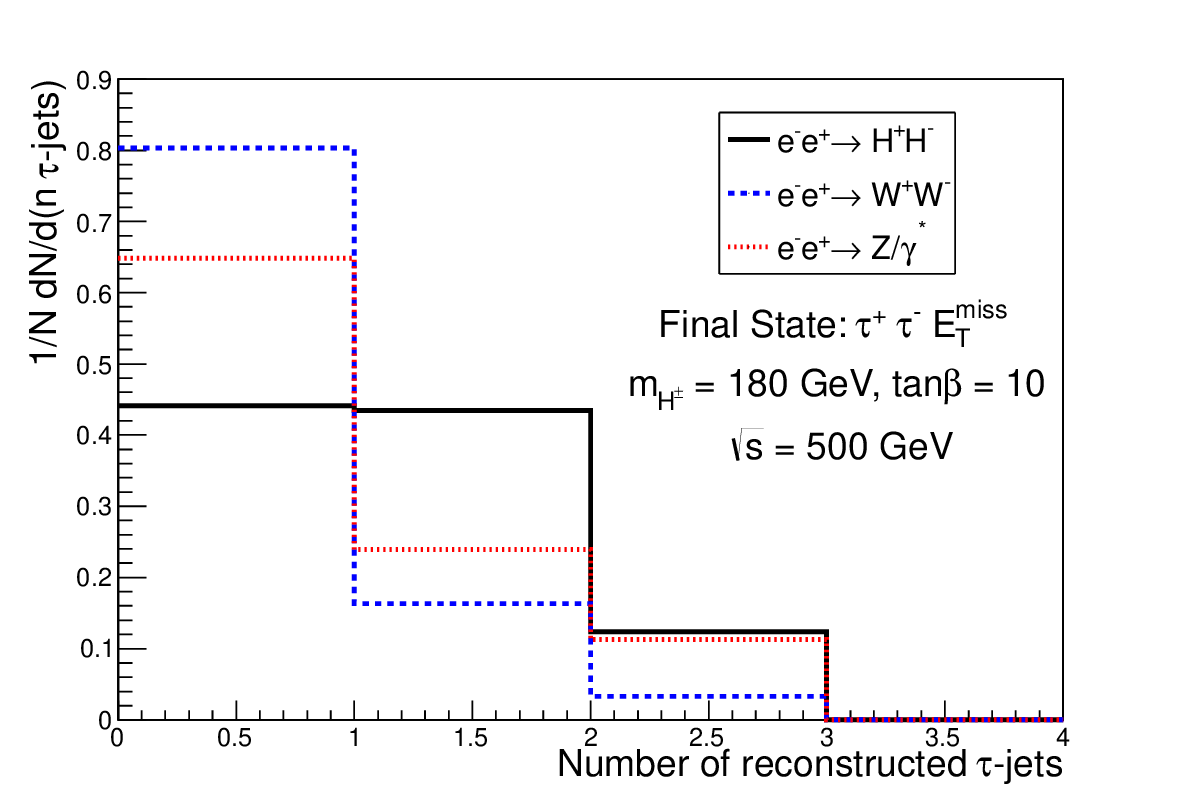}
\end{center}
\caption{The $\tau$ jet multiplicity in signal and background events with center of mass energy of 500 GeV. A $\tau$ lepton is accepted and counted if it passes all selection requirements stated in the $\tau$ identification algorithm description.}
\label{taumul}
\end{figure}

\begin{figure}
\begin{center}
\includegraphics[width=0.6\textwidth]{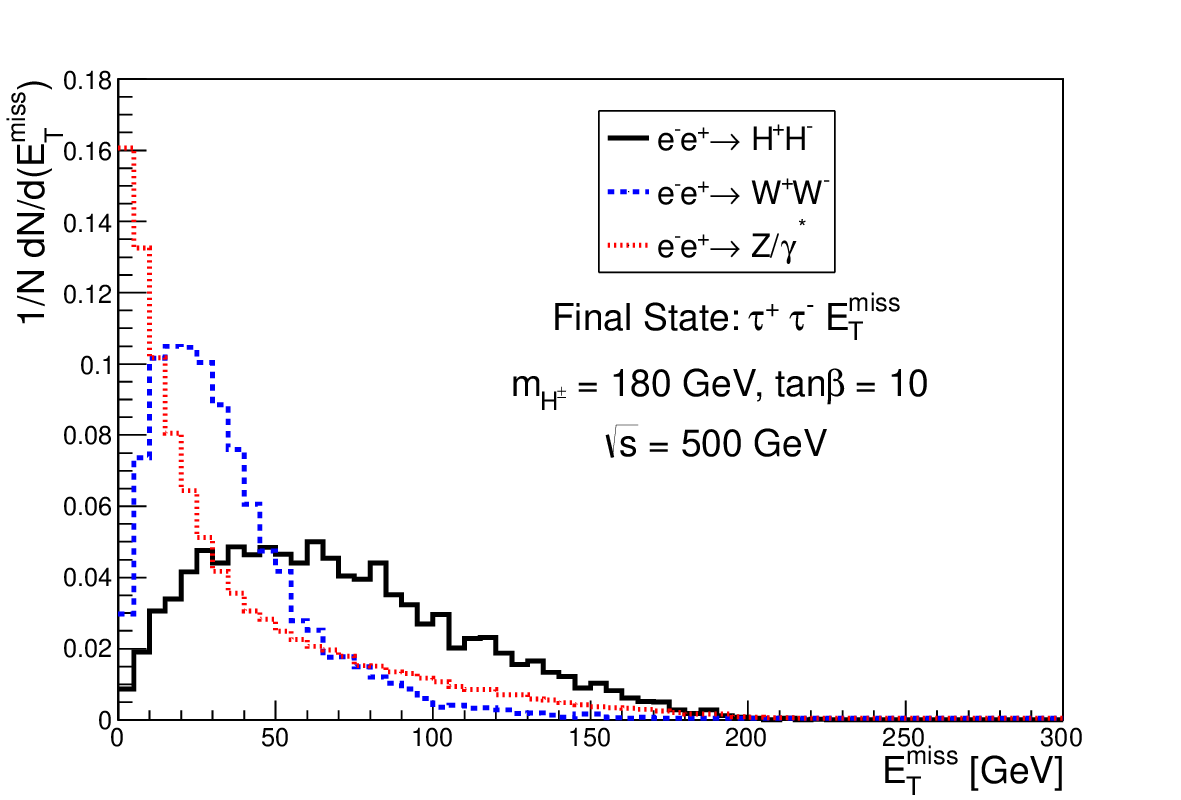}
\end{center}
\caption{The missing transverse energy distribution in signal and background events for a center of mass energy of 500 GeV.}
\label{met}
\end{figure}

In this section, the event selection is described step by step. The analysis strategy is to apply selection cuts inspired by kinematic distributions of the signal and background processes and differences observed in those distributions. The kinematic cut is thus applied to increase the signal to background ratio while keeping the signal at a reasonable level. The signal significance is calculated at the end when total selection efficiencies are obtained for the signal and background.\\
To begin the analysis, when events are generated, jets are reconstructed using FASTJET and sorted in descending $p_{T}$. The missing transverse energy is also calculated as the vectorial sum of transverse momenta of stable particles in the event under the condition that they lie in a pseudorapidity range of $|\eta|<3.5$ with $\eta = -\textnormal{ln}\tan(\theta/2)$. As an example, this angle corresponds to half-angle of the beam pipe cone designed for SiD (Silicon Detector) which will be operating as an ILC detector \cite{sid}. Since there are two $\tau$ leptons and two neutrinos in signal and the background, the two above objects, i.e., jets and MET (missing transverse energy) compose the basic tools for the event analysis.\\ Therefore the event analysis is started by applying the following kinematic thresholds on jets:
\be
E_{T}^{\textnormal{jet}}~>~20~ \textnormal{GeV}~~,~~|\eta|_{\textnormal{jet}}~<~3 
\ee
Figure \ref{njets} shows the jet multiplicity for signal and background events with $\sqrt{s} =$ 500 GeV.
The following condition is applied on every event:
\be
\textnormal{Number~of~jets}~\ge~2
\ee
The selected jets in signal and background events are basically $\tau$-jets produced from the $\tau$ lepton hadronic decays. These jets are characterized by the low charged particle multiplicity (few charged pions from the $\tau$ lepton decay) which results in a narrow jet dominated by an electromagnetic shower. The charged pion energy distributions are correlated with the $\tau$ lepton helicity state which is determined by the spin of the decaying boson. The charged Higgs boson as a spin-less particle decays as the following, $H^{+} \ra \tau^{+}_{L}\nu_{L}$ ($H^{-} \ra \tau_{R}\bar{\nu}_{R}$), whereas the $W$ boson decays like $W^{+} \ra \tau^{+}_{R}\nu_{L}$ ($W^{-} \ra \tau_{L}\bar{\nu}_{R}$). Therefore the $\tau$ leptons have opposite helicity states in two cases. As will be seen later, the one-prong decay is the dominant effect in the $\tau$ lepton hadronic decay. Therefore constraining ourselves to the case of one-prong decay, the $\tau$ leptons from the charged Higgs decay undergo the following decay processes, $\tau^{+}_{L} \ra \pi^{+}\bar{\nu}_{R}$($\tau_{R} \ra \pi^{-}{\nu}_{L}$), whereas in case of W boson decay, the $\tau$ leptons decay as the following, $\tau^{+}_{R} \ra \pi^{+}\bar{\nu}_{R}$($\tau_{L} \ra \pi^{-}{\nu}_{L}$). Therefore in a charged Higgs boson decay to a $\tau$ lepton, the generated $\tau$ lepton, in its decay, tries to push the charged pion forward and kick back the neutrino in order to conserve the angular momentum. The $\tau$ leptons from $W$ boson decays, behave in the opposite way, kicking back the charged pion. As a result the charged pions produced from the charged Higgs boson decays acquire a harder momentum and energy distribution in the laboratory frame. This difference appears as harder jets in signal events compared to background processes. As a conclusion, the signal jets, receive a higher chance to pass the jet kinematic selection and the jet multiplicity distributions look different for the signal and background events as is seen from Fig. \ref{njets}.\\
The selected jets are taken as the $\tau$-jet candidates and are tested with a $\tau$-identification algorithm similar to what is used by LHC experiments \cite{tauid}. The phenomenology of the $\tau$ lepton decay follows studies reported in \cite{tau1,tau2,tau3,tau4,tau5,tau6}. The $\tau$-id starts by considering the fact that the $\tau$ lepton in its hadronic decay produces predominantly one or three charged pions. Due to the low charged track multiplicity in the $\tau$ lepton decay, the charged tracks (pions) in the $\tau$ hadronic decay, acquire relatively a higher transverse momentum compared to tracks of light quark jets. To verify this effect, a jet-track matching cone of $\Delta R = 0.1$ is considered around the jet axis. Here $\Delta R=\sqrt{\Delta\eta^{2}+\Delta\phi^{2}}$ and $\phi$ is the azimuthal angle. Both $\Delta\eta$ and $\Delta\phi$ are calculated for tracks in the jet with respect to the jet axis. The hardest charged track in the matching cone is the candidate for the charged pion from the $\tau$ lepton decay. Figure \ref{ltk} shows distribution of the leading track transverse momentum in signal and background events.
The kinematic cut designed for this distribution is set as the following:
\be
p_{T}^{\textnormal{leading track}}~>~20~ GeV.
\ee
Moreover since $\tau$ jets consist of few charged tracks, they are isolated jets in the tracker. In order to use this fact an isolation cone and a signal cone is defined respectively with cone sizes of $\Delta R < 0.4$ and $\Delta R < 0.07$ around the leading track. The isolation requirement is then as the following:
\be
\textnormal{No charged track with}~ p_{T} > \textnormal{1 GeV in the isolation annulus}~ 0.07 < \Delta R < 0.4
\ee
The low charged track multiplicity in the $\tau$ jets also implies that the leading track in
the jet cone carries a larger fraction of the $\tau$ jet energy compared to quark jets in background events. In order to verify this effect, the distribution of the leading track $p_{T}$ divided by the $\tau$ jet energy is plotted as shown in Fig. \ref{R}.
The applied cut on this distribution is set as the following:
\be
R = p_{T}^{\textnormal{leading track}} / E_{\tau} ~>~ 0.5
\ee
Although harder cuts are also possible, they are avoided here, not to lose the signal statistics.
Finally since the $\tau$ lepton hadronic decay is a one-prong or three-prong decay, the number of charged tracks in the $\tau$ jet is counted by a search in the signal cone. This procedure results in a distribution shown in Fig. \ref{nstk}, according to which the following requirement is applied:
\be
\textnormal{Number of signal tracks in the}~ \tau ~\textnormal{jet}~ =~ 1~ \textnormal{or}~ 3
\ee
It should be noted that in a real situation at the presence of detector smearing, material effects and the magnetic field, a number of signal tracks may fall outside the signal cone. A study of such effects is beyond the scope of this analysis, however, as verified in an LHC experiment study in \cite{LCHCMS}, a very small fraction of signal events appear with two signal tracks and fall in the two-track bin. Therefore these errors are expected to be small. The potential $\tau$ fake rate may be another source of uncertainty for this analysis. A detailed study simulating the linear collider detectors environments is needed to estimate the $\tau$ fake rate, however, the $\tau$ identification efficiency can be tuned with changing the algorithm parameters to have an optimal selection efficiency for real $\tau$'s while keeping the $\tau$ fake rate at a very small value. We are not aware of such a study, however, analyses in CMS \cite{tauidcms} and ATLAS \cite{tauidatlas} collaborations show that with a sophisticated $\tau$ identification algorithm, the real $\tau$ selection can be as high as 70 $\%$ while keeping the fake rate at the level of one percent or so. A linear collider with leptonic beams is expected to have a better performance than LHC due to having a cleaner event environment with smaller particle multiplicity in the event and no underlying event activity which arises in the case of hadronic interactions. Therefore it is expected that $\tau$'s will be under control at a linear collider and the fake rate can be suppressed enough for a reasonable signal selection. The main source of $\tau$ fake rate could be from $Z/\gamma^{*} \ra jj$ background which is suppressed not only by the very low fake rate, but also by the requirement of missing transverse energy threshold. This was checked by running the analysis program and no event survived from $\tau$ tagging and missing transverse energy cuts. The other weak boson pair production backgrounds when in the fully hadronic final state ($WW/ZZ \ra jjjj$) are also suppressed by the same cuts. Figure \ref{taumul} shows the $\tau$ jet multiplicity in signal and background events. 
An event has to have exactly two $\tau$ jets identified by the above algorithm. If this requirement is satisfied, the event is accepted, otherwise it is rejected. \\
In order to reduce the Drell-Yan events, the azimuthal angle between the two $\tau$ jets is calculated.  
\begin{figure}
\begin{center}
\includegraphics[width=0.6\textwidth]{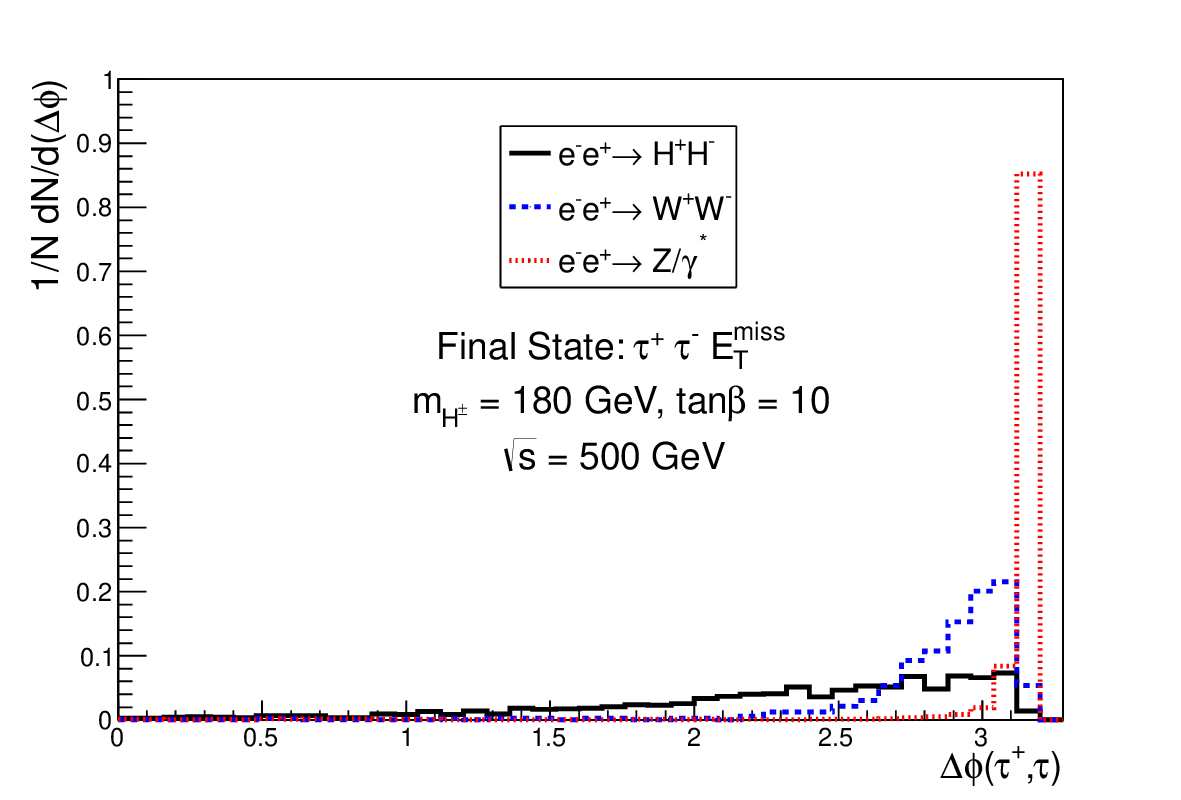}
\end{center}
\caption{The azimuthal angle between the two $\tau$ jets in signal and background events for a center of mass energy of 500 GeV.}
\label{dphi}
\end{figure}
Figure \ref{dphi} shows the distribution of this angle in signal and background events. As is seen, the $Z/\gamma^{*}$ events tend to produce back-to-back $\tau$ jets, as both $\tau$ jets come from the same particle. Therefore the following requirement is applied on each event, 
\be
\Delta\phi_{(\tau^{+},\tau)}~<~3 ~\textnormal{rad}
\ee
The final kinematic distribution to use, is the missing transverse energy which is plotted in Fig. \ref{met}.
The following cut is applied on the missing transverse energy:
\be
\textnormal{Missing transverse energy}~>~30~\textnormal{GeV}
\ee 
Harder cuts are again avoided to keep the signal statistics at a reasonable level. With this cut, the event selection ends and the same procedure is applied on every event and the total selection efficiency is calculated for signal and background samples. This is the topic of the next section.
\section{Results}
\begin{figure}
\begin{center}
\includegraphics[width=0.7\textwidth]{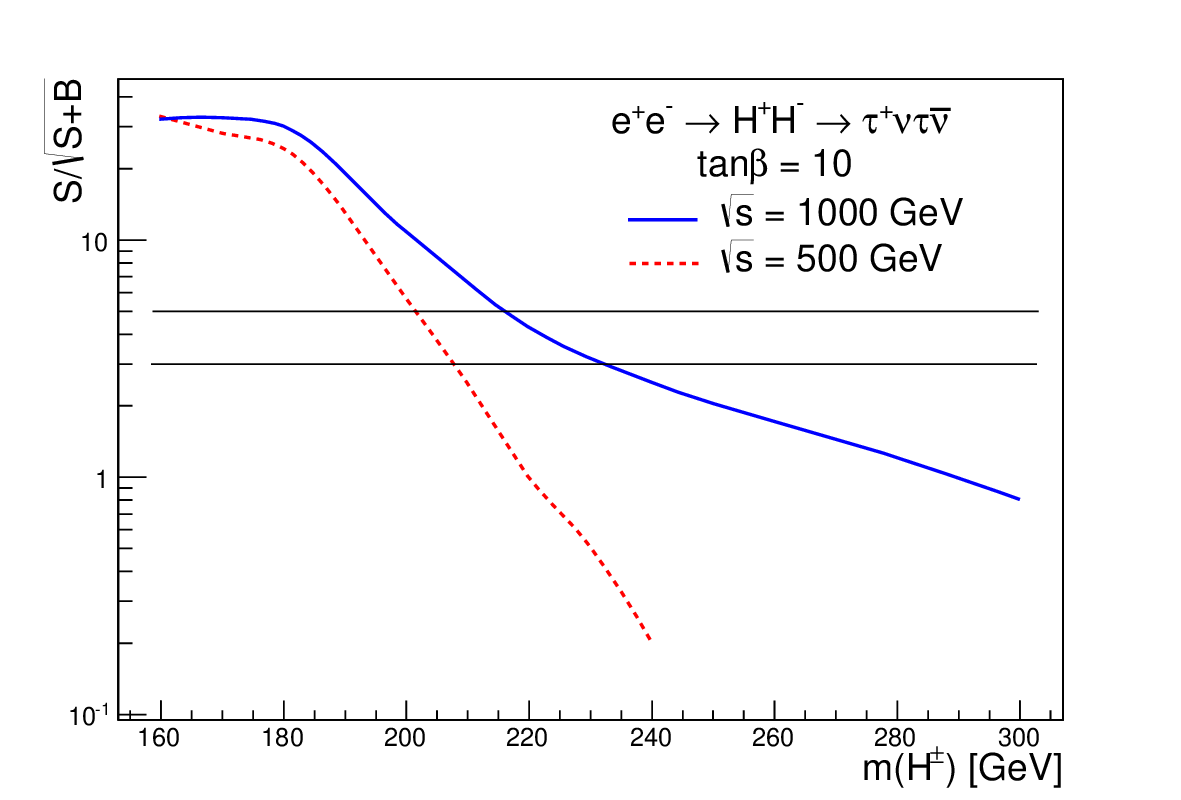}
\end{center}
\caption{The signal significance with $\sqrt{s}=$ 1000(500) GeV shown with solid (dashed) lines with integrated luminosity of 500 $fb^{-1}$.}
\label{sig}
\end{figure}

In this section, selection efficiencies are calculated numerically for different samples of the signal events corresponding to different charged Higgs masses in the most interesting region of the charged Higgs mass spectrum. The \tanb is set to 10 as stated before. Tables \ref{seleff1} and \ref{seleff2} present the results for two center of mass energies of 500 and 1000 GeV respectively. Since for some $m_{H^{\pm}}$ values, the signal is comparable to the background, the signal significance is calculated as in Eq. \ref{signif} where $N_{S}(N_{B})$ is the selected number of signal (background) events. 
\be
\textnormal{Signal Significance}=\frac{N_{S}}{\sqrt{N_{S}+N_{B}}}
\label{signif}
\ee
The sharp drop of the significance in the charged Higgs mass interval of 180-200 GeV is a result of turning on the charged Higgs decay to $t\bar{b}$ which suppresses the $\tau\nu$ decay mode rapidly and starts to be the dominant decay mode for higher charged Higgs masses.\\
Results of Tabs. \ref{seleff1} and \ref{seleff2} are plotted in Fig. \ref{sig}.
\begin{table}
\begin{center}
\begin{tabular}{|c|c|c|c|c|c|c|c|}
\hline
\multicolumn{8}{|c|}{Signal}\\ 
\hline
$m_{H^{\pm}}$ & \tanb & $\sigma(fb)$ & BR($H^{\pm}\ra \tau\nu$) & $\sigma \times \textnormal{BR}^2(H^{\pm}\ra \tau\nu)~(fb)$ & Efficiency & Selected events & Significance \\
\hline
160 & 10 &     51.93 & 0.9773 &    49.6 &     9.73 $\%$ & 2413 & 33 \\
\hline
170&  10&      42.6&  0.9762&      40.6 &     9.88$\%$ &  2006 &  28\\
\hline
180&  10&      36.7&  0.9227&     31.3 &       10.43$\%$ & 1632 & 24\\
\hline
190&  10&      29  &  0.6865&     13.7 &       10.54$\%$ & 722 & 11.9 \\
\hline
200&  10&      22.2&   0.5119&      5.8 &      11.29$\%$ &  327 & 5.7\\
\hline
210&  10&      16.4&  0.3919&      2.52 &      11.08$\%$ &  139  & 2.5\\
\hline
220&  10&      10.1&  0.3181&       1.02&      11.75$\%$ &   60  &  1\\
\hline
230&  10&      5.6&  0.2738 &      0.42 &      12.05$\%$ &  25   &   0.5\\
\hline
240&  10&       2.1&  0.2399&      0.12 &      12.45$\%$ &  8   &   0.2\\
\hline
\multicolumn{8}{|c|}{Background}\\ 
\hline 
\multicolumn{2}{|c|}{Process} & $\sigma(fb)$ & - & $\sigma \times$ BR$~(fb)$ & Efficiency & Selected events & - \\ 
\hline 
\multicolumn{2}{|c|}{$Z/\gamma^{*}$} & 16700 & - & 1030 & 0.45$\%$ & 2250 & - \\
\hline
\multicolumn{2}{|c|}{$W^{+}W^{-}$} & 7603 & - & 90.6 & 1.46$\%$ & 661 & - \\
\hline
\multicolumn{2}{|c|}{$ZZ$} & 451 & - & 15 & 0.48$\%$ & 36 & - \\
\hline 
\end{tabular}
\end{center}
\caption{Signal and background cross sections times branching ratios and their selection efficiencies. The last column is the signal statistical significance obtained from the final selected number of events which are normalized to a total integrated luminosity of 500 $fb^{-1}$. The center of mass energy of the collider is set to 500 GeV. \label{seleff1}}
\end{table}
\begin{table}
\begin{center}
\begin{tabular}{|c|c|c|c|c|c|c|c|}
\hline
\multicolumn{8}{|c|}{Signal}\\ 
\hline
$m_{H^{\pm}}$ & \tanb & $\sigma(fb)$ & BR($H^{\pm}\ra \tau\nu$) & $\sigma \times \textnormal{BR}^2(H^{\pm}\ra \tau\nu)~(fb)$ & Efficiency & Selected events & Significance \\
\hline
160& 10& 27.7&      0.9773&  26.5&        14.1$\%$&   1865&   32\\
\hline
170& 10& 26.5&      0.9762&  25.2 &       15.1$\%$&   1906&   32\\
\hline
180& 10& 25.7&      0.9227&  21.9&        15.6$\%$&   1707&   30\\
\hline
200& 10& 24.1&      0.5119&  6.31 &       16.1$\%$&   508 &   11\\
\hline
220& 10& 22  &      0.3181&  2.23 &       16.1$\%$&   180 &   4.3\\
\hline
240& 10& 20.6&      0.2399&  1.2 &        17.1$\%$&   103 &   2.5\\
\hline
260& 10& 18.7&      0.1999&  0.75 &       17.8$\%$&   67  &   1.7\\
\hline
280& 10& 16.7&      0.1759&  0.52 &       18.1$\%$&   47  &   1.2\\
\hline
300& 10& 14.8&      0.1605&  0.38 &       18.7$\%$&   36  &   0.8\\
\hline
\multicolumn{8}{|c|}{Background}\\ 
\hline 
\multicolumn{2}{|c|}{Process} & $\sigma(fb)$ & - & $\sigma \times$ BR$~(fb)$ & Efficiency & Selected events & - \\ 
\hline 
\multicolumn{2}{|c|}{$Z/\gamma^{*}$} & 4307 & - & 271 & 0.92$\%$ & 1247 & - \\
\hline
\multicolumn{2}{|c|}{$W^{+}W^{-}$} & 3179 & - & 37.4 & 1.42$\%$ & 265 & - \\
\hline
\multicolumn{2}{|c|}{$ZZ$} & 175 & - & 5.8 & 0.92$\%$ & 27 & - \\
\hline 
\end{tabular}
\end{center}
\caption{Signal and background cross sections times branching ratios and their selection efficiencies. The last column is the signal statistical significance obtained from the final selected number of events which are normalized to a total integrated luminosity of 500 $fb^{-1}$. The center of mass energy of the collider is set to 1000 GeV. \label{seleff2}}
\end{table}

\section{Comparison of the Results with Previous Studies}
In this analysis, the $\tau^-\tau^+ E^{miss}_T$ final state was analyzed as a search channel for a heavy charged Higgs boson. Let us call it ``Analysis B''. A previous study reported in \cite{sch6,sch7} analyzed the hadronic decay of the charged Higgs boson, i.e., $H^{\pm} \ra t\bar{b}$ while the other (off-shell) charged Higgs decays to a $\tau\nu$ pair. Let us call that analysis, ``Analysis A''. There are different aspects of the two analyses which can be used to compare them.\\
\textbf{Analysis A}:\\
The final state in this analysis, contains two light jets, two b-jets, and a $\tau$ jet. Since in total there are five jets in the final state, a reasonable understanding of the jet reconstruction algorithm, the jet energy scale uncertainty, the b-tagging efficiency and b-jet mistagging rate is needed to assess the signal observability through this final state.\\
The charged Higgs transverse mass can be reconstructed in this final state and the event rate is more than that in analysis B due to the larger charged Higgs branching ratio of decay to $t\bar{b}$ as compared to $\tau\nu$, however, a full event selection relies on reconstructing five jets and applying $\tau$ ID and b-tagging on them. These algorithms and the kinematic cuts applied on $p_T$ and $|\eta|$ of the jets in the event, are subject to uncertainties which could be high. A study of such uncertainties needs a full detector simulation including jet activities in the detector material, fake jets as a result of electronic noise, etc. Since the number of required jets for the full event selection is high, finite uncertainties of the types mentioned above could arise large uncertainties in the final event selection and conclusions.\\
\textbf{Analysis B}:\\
The analysis presented in this paper, relies only on $\tau$ jets and missing transverse energy. The event final state has a low physical object multiplicity (only two $\tau$'s and $E^{miss}_T$) and the only uncertainties in the event selection are related to the $\tau$ ID and $E^{miss}_T$ estimation algorithms. Although the heavy charged Higgs has a smaller branching ratio of decay to $\tau\nu$ as compared with $t\bar{b}$, this final state selection may be more reliable than the hadronic one. The $\tau$ ID presented in this work was based on a simple cut-based selection, however, it can be improved using complicated algorithms like those currently being used at LHC \cite{tauidcms,tauidatlas}. These algorithms provide efficiencies more than 70$\%$ at LHC while keeping the fake rate negligible. A similar $\tau$ ID algorithm applied at a linear collider environment is expected to perform even better due to the less event contamination in leptonic collisions and it is expected that results of this analysis are improved when using state-of-the-art $\tau$ ID algorithms in the event selection. Therefore the problem of smaller branching ratio of charged Higgs decay to $\tau\nu$ can be compensated by using sophisticated $\tau$ ID algorithms with high efficiencies and low fake rates.\\
The aforementioned concerns on the uncertainties related to the jet reconstruction and b-tagging may be serious as an experience with CMS analyses already showed it. The analysis reported in \cite{LowetteCH} concluded that no sizable region of the parameter space is left with an observable heavy charged Higgs signal in the hadronic final state with reasonable assumptions on the jet and b-jet uncertainties, while in \cite{HCHCMS}, a heavy charged Higgs signal turned out to be observable through its decay to $\tau\nu$. The final conclusion at LHC has thus been to use the charged Higgs decay to $\tau\nu$ as the main search channel in both light and heavy regions although the branching ratio of decay to this particular channel reduces in the heavy charged Higgs region.\\
The analysis presented in this work provides an alternative way of searching for the heavy charged Higgs in addition to analysis A. Both analyses try to estimate a future linear collider potential for a heavy charged Higgs observation below the kinematic threshold. It is early to judge which channel provides the potential to successfully observe the charged Higgs signal without a knowledge of the detector uncertainties, however, both final states can be used independently as a search channel for this particle at a linear collider. Results can of course be compared and/or combined to achieve a better statistical significance. The analysis A can be used for the charged Higgs transverse mass reconstruction and obtaining an estimate on the charged Higgs mass, while analysis B can be considered as a supporting analysis of the ``counting'' type which looks for an excess of events over SM. They are complementary and useful in providing independent sources of the heavy charged Higgs boson. Therefore it is reasonable to perform both of them for interpreting the final results.
\section{Conclusions}
The charged Higgs pair production in a linear $e^{+}e^{-}$ collider was studied looking at the $\tau$ lepton pair final state. Results show that the charged Higgs is observable through this channel for a wide range of the charged Higgs mass. With $\sqrt{s}$ = 500 GeV, having collected data corresponding to an integrated luminosity of 500 \invfb, the signal is observable with $5\sigma$ significance up to $m_{H^{\pm}}\simeq$ 200 GeV. The observability of the signal turns out to be extended up to $m_{H^{\pm}}\simeq$ 220 GeV with $\sqrt{s}$ = 1000 GeV. The reason that no dramatic extension of the $5\sigma$ contour is obtained when increasing the center of mass energy of the collider could be the fact that the signal statistics remains low for heavy charged Higgs masses. This is a reflection of two facts. First, being a lepton collider, the overall cross sections involved in this analysis decrease when increasing the center of mass energy of the collider. This experience is contrary to the case of hadron colliders where the total cross section of such events normally increase when the center of mass energy is increased. Second, the branching ratio of charged Higgs decay to $\tau\nu$ slows down to less than 0.2 in the high mass region while the charged Higgs decay to $t\bar{b}$ is high enough to produce the main signal in that region. Nevertheless the signal studied in this work serves as a detectable signal in a large area of parameter space and can be used as a complementary search to those already proposed.

\end{document}